ARTICLE

# Flash Freeze–Thaw Phenomenon in Sprayed Evaporating Micrometer Droplets

Junshi Wang[a], Zehao Pan[a,b], Howard A. Stone[a,b], and Maksim Mezhericher[a,b]

[a]Department of Mechanical and Aerospace Engineering, Princeton University, Princeton, New Jersey 08544, USA
[b]Inaedis, Inc., Princeton, New Jersey 08540, USA

ABSTRACT
Two-fluid spray nozzles are widely used in combustion, chemical processing, pharmaceutical coating, environmental control, and spray drying to atomize liquids with pressurized gas. However, the adiabatic cooling and resulting *flash freeze–thaw* exposure of atomized droplets remain underexplored. Using high-fidelity computational fluid dynamics coupled with droplet-scale nucleation modeling, we show that the atomizing gas temperature at the nozzle exit can fall from 22 °C to below −130 °C, initiating rapid ice nucleation and freezing in micron-scale droplets. For atomizing gas at 5 bar (gauge) and 22 °C, all droplets smaller than 1.5 $\mu$m freeze, whereas droplets larger than 3 $\mu$m remain liquid. These frozen droplets thaw within $\mathcal{O}(10)$ $\mu$s upon leaving the cold zone, subjecting sensitive actives to intense freeze–thaw thermomechanical stresses near the nozzle even when the bulk drying gas is warm. Parametric studies show that ice formation is eliminated at atomizing gas temperatures above 110 °C for all gas-to-liquid mass ratios (GLRs) between 8 and 25, or at GLR < 12 for all atomizing gas temperatures; the chamber drying gas does not influence near-nozzle freezing. Additionally, we demonstrate that swirling flow intensifies flash freeze–thaw by deepening gas cooling, whereas non-swirling flow extends cold-zone residence time, yet both designs produce similar iced-droplet fractions. We construct an operating map delineating conditions that avoid flash freeze–thaw and show that the no-ice boundary provides a conservative criterion for both swirl and non-swirl nozzles. These findings identify a previously unrecognized freeze–thaw stress mechanism that can compromise spray-dried pharmaceutical product stability.

## 1. Introduction

Two-fluid nozzles are used extensively in spray drying in the pharmaceutical and food industries to produce atomized liquid droplets, which are then dried into powders. Two fluids are involved in such liquid atomization process: pressurized gas flows through a nozzle to atomize a feedstock liquid into small droplets.

When the pressurized atomizing gas is released from the nozzle exit to a lower-pressure environment, adiabatic expansion can cause an abrupt decrease in gas temperature and density, forming a low-temperature region downstream of the nozzle.

---

For example, in a two-fluid nozzle used by a typical pilot-scale spray drying facility, GEA Niro Mobile Minor (*1*), the atomizing gas pressure can be as high as 5 bar above atmospheric pressure, which can cause a temperature drop of over 100 K based on an ideal adiabatic expansion process.

The low-temperature region may cause substantial cooling, and even freezing, of the liquid droplets passing through it. Subsequently, when iced droplets leave the cool region, they will warm up and thaw. The cooling process and potential freeze–thaw cycles of liquid droplets generate thermal stresses, which can degrade products like proteins in the liquid that are sensitive to such conditions. Our hypothesis of the potential impact of rapid freeze–thaw cycles can be partly supported by the flash freezing of supercooled droplets (*2*), where researchers have observed freezing of 10-micron-sized droplets in several microseconds.

To our knowledge, neither the adiabatic cooling of atomizing gas exiting a two-fluid nozzle nor the resulting droplet freezing has been studied systematically for common experimental conditions, and no experimental data are available to characterize gas cooling in such nozzles. Existing numerical work suggests the presence of a low-temperature region near the nozzle (*3, 4*), but the temperature field has not been resolved with sufficient detail to determine its spatial extent, magnitude, or impact. More broadly, despite extensive research on spray-drying hydrodynamics and dryer-scale modeling (*5–10*), the near-nozzle adiabatic expansion, its impact on local gas temperature fields, and the resulting probability of droplet freezing have not been quantified for realistic industrial operating conditions. Consequently, it is still unclear which droplet size classes are most susceptible to freezing and on what time scales.

We combine computational fluid dynamics (CFD) simulations and analytical modeling to investigate the flow through a two-fluid nozzle. Specifically, we document the adiabatic cooling of the atomizing gas owing to expansion at the nozzle exit and, correspondingly, the cooling and freezing of the sprayed droplets. We created a 3D full-scale computational model based on the GEA Niro Mobile Minor spray dryer (*1*), with the nozzle modeled based on a Schlick two-fluid nozzle (*11*), and conducted computational fluid dynamics (CFD) simulations for the spray drying process. We built an analytical model to calculate ice nucleation, both for homogeneous and heterogeneous types, and growth inside supercooled droplets, and compared the percentage of iced droplets, both in number and volume, between nozzles with swirl and without swirl of the gas phase. Further, we conducted parametric studies on the gas-to-liquid mass ratio (GLR) and the temperature of atomizing gas, and summarized our results in an operating map to identify working conditions of the two-fluid nozzle to avoid ice formation in droplets.

## 2. Materials and Methods

### 2.1. *Ice nucleation and growth in supercooled water*

Homogeneous ice nucleation occurs when the temperature of supercooled water is below 240 K. We adopt the nucleation rate in the literature (*12*) (Figure 1a). Heterogeneous ice nucleation can happen at higher temperatures, and the nucleation rate varies with the types of impurities. Here we adopt classical theory for heterogeneous nucleation supported by numerical modeling work (*13*) (Figure 1a). The heterogeneous nucleation rate, $J_{\text{het}}(T)$, is modeled using an analytical expression derived from fitting experimental data,

$$J_{\text{het}}(T) = \exp\left(\ln(A_{\text{het}}) + \frac{C_{\text{het}}}{(T - T_m)^2 T}\right), \quad (1)$$

where $A_{\text{het}}$ is the fitted kinetic prefactor for heterogeneous nucleation and thus has the same units as $J_{\text{het}}(T)$ (m$^{-3}$ s$^{-1}$). The fit yields $\ln(A_{\text{het}}) = 102 \pm 7.7$ and $C_{\text{het}} = -(1.7 \pm 0.07) \times 10^7$ K$^3$ (13), where $C_{\text{het}}$ is a fitting constant and $T_m$ is the equilibrium melting temperature of ice in the monatomic water model ($T_m = 274.6$ K) (14). Note that the nucleation rate units in Figure 1a are converted from (m$^{-3}$ s$^{-1}$) to ($\mu$m$^{-3}$ $\mu$s$^{-1}$) for consistency with the length and time scales relevant to the flash freeze–thaw phenomenon of micrometer water droplets.

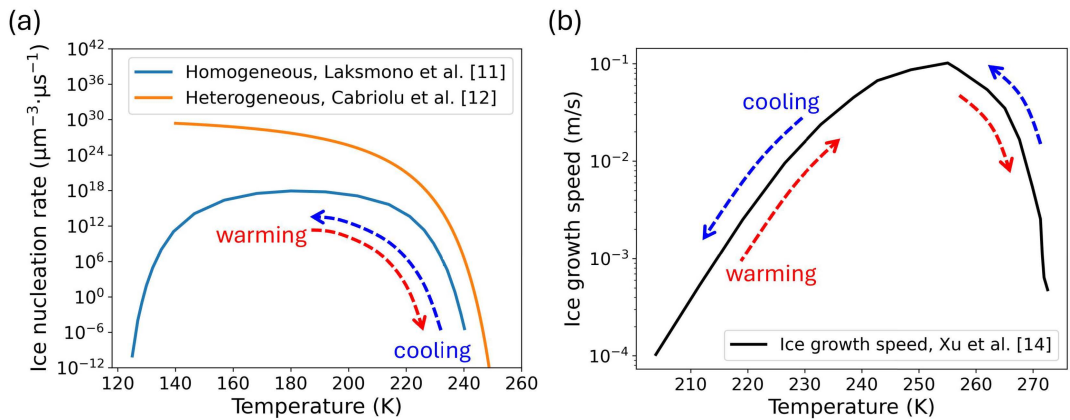

**Figure 1.** (a) Ice nucleation rates adopted from the literature for homogeneous nucleation (12) and heterogeneous nucleation (13), respectively. (b) Ice growth rate in supercooled water adopted from the literature (15). Dashed blue arrows indicate cooling; dashed red arrows indicate warming.

We assume ice grows uniformly along the radius of the nucleus once it is formed. We adopt the speed of ice growth in supercooled water reported in the literature (15) (Figure 1b). While the nucleation parameters are not specific to any one solute or impurity type, the resulting ice fractions and time scales are expected to be representative of the order of magnitude of freeze–thaw cycles, rather than exact predictions for a specific formulation.

## 2.2. *Modeling of single droplet evaporation*

To calculate the temperature for droplets of varying diameters, we employed our previously developed analytical model of single droplet evaporation kinetics (16–19). The model assumes a spatially uniform droplet temperature, $T_d(t)$, justified by the lumped capacitance assumption (20). This requires that the Biot number (Bi = $hL_c/k_d$) be small ($\ll 1$), where $h$ is the convective heat transfer coefficient, $k_d$ is the thermal conductivity of the droplet, $L_c = V_d/A_d = R/3$ is the characteristic length, with $V_d$ the droplet volume and $A_d$ the surface area. The time required for thermal equilibration is quantified by the Fourier number (Fo), defined as Fo = $\alpha_d t/R^2$, where $\alpha_d$ is the thermal diffusivity, $t$ is time, and $R$ is the droplet radius. We calculated Bi and Fo for droplets with diameters $D < 10\,\mu$m. We find that Bi $< 0.1$ and Fo $> 1$, confirming both the uniform temperature assumption and rapid thermal equilibration with the environment. Consequently, the energy balance for an evaporating droplet in drying gas flow can be approximated by (*19*):

$$m_d c_p \frac{dT_d}{dt} = h(T_g - T_d)A_d - \dot{m}_v h_{fg}, \qquad (2)$$

where $h_{fg}$ is latent heat of vaporization, $\dot{m}_v$ is mass flow rate of vapor from the droplet, $c_p$ is specific heat capacity of droplet, $m_d$ is droplet mass, $h$ is the convective heat transfer coefficient, $A_d$ is the droplet surface area, with $A_d = 4\pi R^2$ and $R$ the droplet radius, $t$ is time, and $T_g$ and $T_d$ are temperatures of the drying gas and the droplet, respectively. The change in the droplet radius due to evaporation is calculated according to (*19*):

$$\frac{dR}{dt} = -\frac{\dot{m}_v}{4\pi R^2 \rho_l}, \qquad (3)$$

where $\rho_l$ is the density of the liquid fraction in the evaporating droplet. The mass flow rate from droplet evaporation is given by (*19*):

$$\dot{m}_v = h_D \left(\rho_{v,s} - \rho_{v,\infty}\right) 4\pi R^2, \qquad (4)$$

where $h_D$ is the convective mass transfer coefficient, and $\rho_{v,s}$ and $\rho_{v,\infty}$ are partial vapor densities at the droplet surface and far away from the droplet. To evaluate the coefficients of heat and mass transfer, $h$ and $h_D$, we use Nusselt and Sherwood numbers determined from the Ranz-Marshall correlations (*16*):

$$\mathrm{Nu}_d = \frac{2hR}{k_f} = 2 + 0.6\,\mathrm{Re}_d^{1/2}\,\mathrm{Pr}^{1/3} \qquad (5)$$

$$\mathrm{Sh}_d = \frac{2h_D R}{D_v} = 2 + 0.6\,\mathrm{Re}_d^{1/2}\,\mathrm{Sc}^{1/3}. \qquad (6)$$

Here $\mathrm{Nu}_d$ and $\mathrm{Sh}_d$ are droplet Nusselt and Sherwood numbers, $k_f$ is thermal conductivity of a gas-vapor film adjacent to the droplet, $D_v$ is mass diffusivity of the vapor, $\mathrm{Re}_d = \frac{2v_{rel}R}{\nu_g}$ is the droplet Reynolds number for motion in air with $v_{rel}$ being the relative velocity between gas and droplet, $\nu_g$ is the kinematic viscosity of the gas, and Pr and Sc are, respectively, Prandtl and Schmidt numbers of the drying gas.

### 2.3. *Modeling of nucleation events and ice growth in droplets*

We consider an ice nucleation event that happens once the droplet temperature falls below the threshold for homogeneous and heterogeneous nucleation. We assume these two types of nucleation are independent events within each droplet. The number of new ice nuclei, $\Delta N_i$, formed during the time interval $[t_i, t_{i+1}]$ is calculated as:

$$\Delta N_i = \lfloor J(T_d(t_i)) \cdot V_d(t_i) \cdot \Delta t \rfloor, \qquad (7)$$

where $J(T_d(t_i))$ is the nucleation rate at droplet temperature $T_d(t_i)$, $V_d(t_i)$ is the droplet volume at time $t_i$, and $\Delta t$ is the time step. The floor function, $\lfloor \cdot \rfloor$, is used because the number of nuclei must be an integer.

We assume that once a nucleus is formed, it grows as a spherical ice crystal, with radius $r(t)$. The radius of a single ice crystal that nucleated at time $t_\text{nuc}$ and is observed at a later time $t_i$ is given by integrating the growth speed $v(T_d(t'))$(Figure 1b). In the discrete simulation, this integral is approximated by a sum:

$$r(t_i, t_\text{nuc}) = \int_{t_\text{nuc}}^{t_i} v(T_d(t'))dt' \approx \sum_{j=j_\text{nuc}}^{i} v(T_d(t_j)) \cdot \Delta t. \tag{8}$$

The volume of this single spherical crystal is therefore:

$$V_\text{crystal}(t_i, t_\text{nuc}) = \frac{4}{3}\pi \left[r(t_i, t_\text{nuc})\right]^3. \tag{9}$$

The total ice volume $V_{ice}(t_i)$ at time $t_i$ is the sum of the volumes of all crystals that have nucleated up to that point. This is calculated by summing over all nuclei formed at previous time steps:

$$V_\text{ice}(t_i) = \sum_{j=0}^{i} \left(\Delta N_j \cdot V_\text{crystal}(t_i, t_j)\right). \tag{10}$$

The total ice volume is the sum of contributions from homogeneous nucleation and heterogeneous nucleation and is physically constrained by the total volume of the droplet.

## 2.4. *Multiphase flow modeling using Computational Fluid Dynamics*

The numerical modeling and simulations were performed using the ANSYS Fluent 2025 R2 software package, building upon the CFD model of the spray drying process previously developed by Mezhericher et al. (*17*, *18*). Specifically, we conducted full-scale, three-dimensional, steady-state CFD simulations of the spray drying process (see Figure 2) using a computational model of the GEA Niro Mobile Minor spray dryer (*1*), in which the nozzle was modeled after a Schlick two-fluid nozzle (*11*). The gas phase was modeled as a mixture of nitrogen and water vapor following ideal gas laws and mixing laws; the liquid droplets were modeled as water with a 7.5 wt% non-evaporating solid component which has the same properties as water. The total height of the inner volume of the drying chamber (see Figure 2a) is 1315 mm with the straight wall height of 620 mm and the converging wall height of 590 mm at a 60° angle. The drying gas inlet is elevated for 25 mm, and the outlet is extended for 80 mm. The inner diameter of the chamber is 800 mm. The drying gas inlet is an annulus with 150 mm outer diameter and 50 mm inner diameter. The inner diameter of the outlet is 120 mm.

In this study, we investigated two types of gas-liquid mixing in the two-fluid nozzle: swirling and non-swirling atomizing gas flow. Initially, we utilized the unsteady flow solver in the Fluent CFD package to investigate the swirl nozzle (Figure 2e) and non-swirl nozzle (Figure 2f) configurations, both at various atomizing gas pressure and atomizing gas temperature. These computations demonstrated that the flow regime in all configurations converged to a steady-state flow. Therefore, to reduce the computational burden, we subsequently adopted a steady-state flow solution in all further numerical simulations. An Eulerian-Lagrangian model was employed to simulate the multiphase flow, treating the gas flow as a continuous phase (Eulerian approach) and droplets/particles as the discrete phase (Lagrangian approach). The gas phase was governed by the Reynolds-Averaged Navier-Stokes (RANS) equations and the energy conservation equations, incorporating gravity and ideal gas behavior. The viscous modeling utilized the SST $k$-$\omega$ turbulence model. Species transport is considered a nitrogen-water vapor mixture as an ideal gas. The spray of evaporating droplets was modeled using the discrete phase model (DPM) by incorporating multi-component droplet injections at the nozzle outlet (Figure 2a).

We modeled the gas-liquid mixing at the nozzle outlet, but not the atomization process in detail. In our simulations, all droplets are released 3 mm downstream of the nozzle exit with a velocity magnitude of 0.1 m/s, using a solid-cone injection type in the DPM model with an outer injection radius of 1 mm. For all simulations, the mass flow rate of liquid is fixed at 0.02 kg/min (20 mL/min). A total of 1000 streams of injections were adopted. Note that the 1000 droplets shown in the simulations hereafter represent computational parcels, not individual physical droplets. Each parcel corresponds to a large number of identical droplets with the same size, velocity, and thermodynamic properties, and is used to statistically represent the spray population. For each injection, the droplet size distribution of injected droplet parcels in the ANSYS Fluent numerical model follows the measured droplet size distribution function, which is described in detail in the following Section 2.5.

Specifically, the droplet size distribution of injected droplet parcels in the DPM model was given by a user-defined function (UDF) in the tabulated form of mass fractions, based on Section 2.5 and Figure 4, and the UDF is provided in the Supplementary Information. In the droplet size distribution used, the minimum droplet diameter is 0.9 μm, the maximum droplet diameter is 86 μm, the volume median diameter, Dv50, is 21.6 μm, and the number median diameter, Dn50, is 1.4 μm.

Droplet drying kinetics were simulated using a two-component mixture model: evaporating water and non-evaporating solid material. For simplicity, both components were assigned water's density, thermal conductivity, and heat capacity. Droplets with water content below 5 wt% were treated as non-evaporating spherical particles in subsequent calculations. As droplets traversed the spray drying chamber, carried by the mixed streams of the drying and atomizing gases, they were considered either "escaped" when exiting the outlet, or "trapped" when deposited on the chamber walls.

The boundary conditions included mass flow inlets for drying gas (Figure 2c) introduced via an annulus chamber inlet and pressure inlet for atomizing gas (Figure 2d) injected at the annulus nozzle inlet. Both drying and atomizing gases were set as pure nitrogen. The drying gas flow rate was set at 6 kg/min (4801 SLPM) and drying gas temperature at 22 °C. We performed parametric studies on the flow rate and temperature of the atomizing gas. To study the effect of atomizing gas flow rate, we set the atomizing gas inlet gauge pressure at one of five values: 1.2 bar, 2 bar, 3 bar, 4 bar, and 5 bar, resulting in the atomizing gas flow rates 0.17 kg/min, 0.25 kg/min, 0.34 kg/min, 0.42 kg/min, and 0.5 kg/min, respectively. Therefore, the corresponding gas-to-liquid mass ratios (GLRs) were 8.4, 12.5, 16.8, 21.0, and 25.1. To study the effect of atomizing gas temperature, we set the atomizing gas temperature at one of five values: 22 °C (295 K), 50 °C (323 K), 80 °C (353 K), 110 °C (383 K), and 140 °C (413 K). A pressure outlet was applied at the chamber exit (Figure 2b). The chamber wall was modeled as a no-slip thermally conducting surface with convective thermal conditions, assuming 22 °C outside ambient temperature.

The numerical computational domain utilized a polyhedral mesh (Figure 3a) with local mesh refinement in regions downstream of the drying gas and atomizing gas in-lets, where the velocity magnitude gradients were high, to capture the flow features with greater accuracy (Figure 3a-c). 20 layers of inflation mesh were used on the walls to ensure a no-slip boundary condition and that the maximum y-plus value of all wall boundaries is around 1. Mesh-size studies were performed to ensure numerical stability, solution accuracy, and mesh independence, following the same grid-independence procedure described in our previous work (*21*). Based on these studies, the final simulations were carried out using a volume mesh with 5.5 million mesh cells for the geometry; details of the assessment of the mesh-independence and the selection of the mesh size and geometry are provided in (*21*). The criteria for a converged solution were defined as residuals reaching $10^{-3}$ for continuity, momentum, $k$, and $\epsilon$, and $10^{-6}$ for energy and species. Typically, convergence was achieved after 1000 numerical iterations. The computations were performed on clusters of Princeton Research Computing. A typical simulation took approximately 8 hours, utilizing 16 parallel 2.6 GHz Intel Skylake CPU cores.

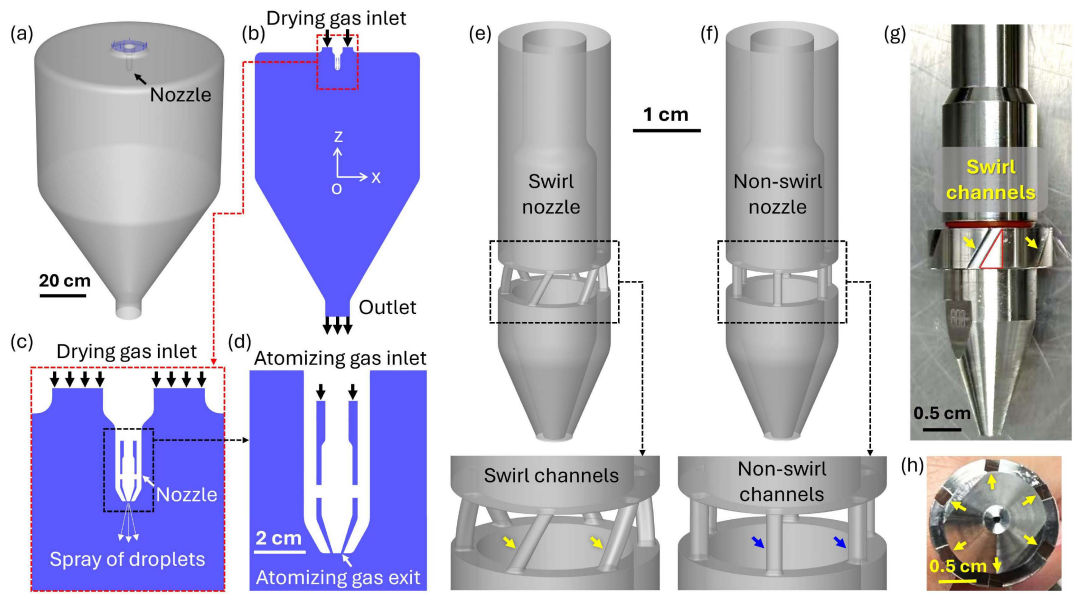

**Figure 2.** Computational modeling of the spray drying process. (a) Drying chamber and spray nozzle geometry. (b) Vertical cross section of the three-dimensional computational fluid domain. (c) Magnified view of the drying gas inlet, nozzle, and the location of cone-shaped droplet injections from the spray nozzle. (d) Magnified view of the atomizing gas inlet and exit. (e) Nozzle geometry with six twisted channels (yellow arrows) to create swirling flow in the atomizing gas. (f) Nozzle geometry with six straight channels (blue arrows) to create non-swirling flow in the atomizing gas. (g) Photograph of the central component of a real two-fluid nozzle with six swirl channels of 30° twisting angle. (h) Photograph of the bottom view of the central nozzle component showing the outlets of six swirl channels uniformly spaced around the axis.

### 2.5. *Measurement of droplet size distribution*

We measured the droplet size distribution of the sprayed water droplets from a standard two-fluid nozzle of the BUCHI Mini Spray Dryer B-290 at a feedstock liquid flow rate of 30 mL/min under an atomizing gas pressure of 2 bar. The measured distribution (Figure 4) is used as input for the DPM injections in all subsequent simulations, so that the effect of atomizing gas pressure or temperature can be studied independently of droplet size.

The droplet size distribution of the atomizer was measured with laser diffraction

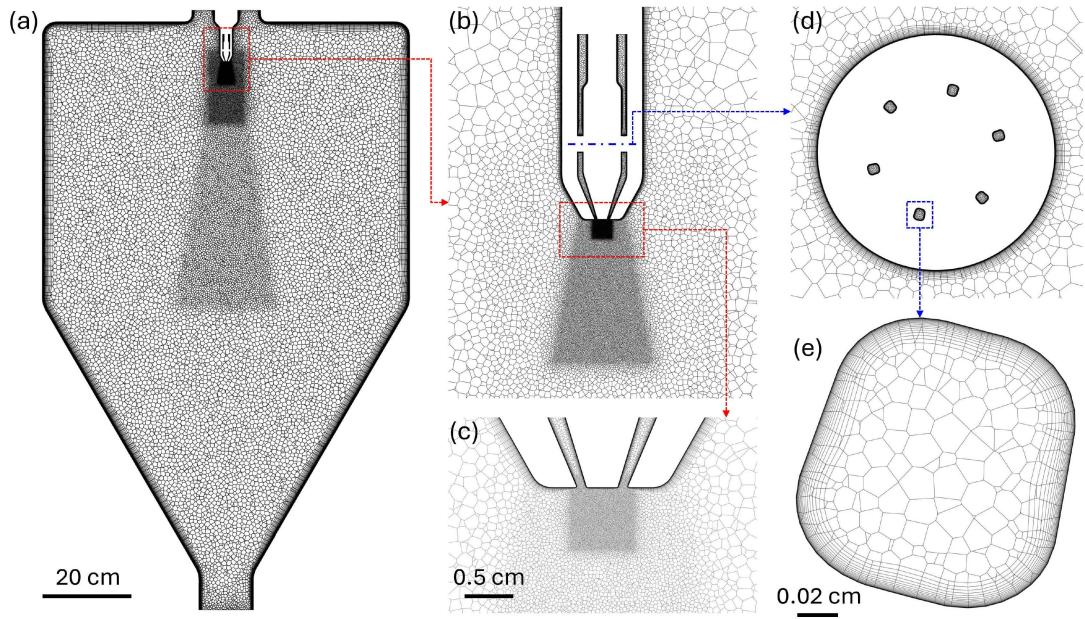

**Figure 3.** Numerical mesh of the computational model. (a) Polyhedral fluid mesh on the central vertical cross section of the computational domain. (b) Zoomed-in view of the mesh surrounding and downstream of the nozzle. (c) Further magnified view of the nozzle exit. (d) Mesh on the horizontal cross section of the nozzle cutting through the six swirl channels. (e) Zoomed-in view of a single channel. Local mesh refinement was applied to improve resolution.

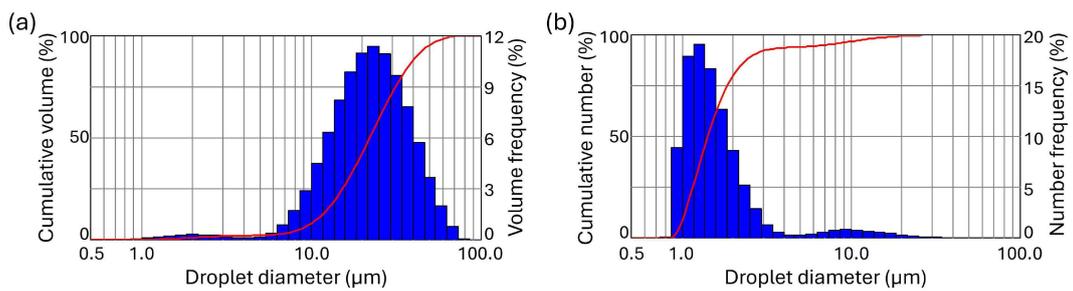

**Figure 4.** Droplet size distribution measurements of sprayed water droplets from the two-fluid nozzle of the BUCHI Mini Spray Dryer B-290 as the input for DPM injections in all numerical simulations of the present study. The results are obtained at a feedstock liquid flow rate of 30 mL/min under an atomizing gas pressure of 2 bar. The droplet size measurements are expressed as histograms of (a) volume-based distribution and (b) number-based distribution.

(Spraytec, Malvern Panalytical, UK) using the previously described procedure (*22*). The nominal measurement accuracy of the Malvern Spraytec device is better than ±1% for the median diameter of a droplet size distribution. For the experimental run to measure the droplet size distribution, aerosol was generated for a minimum of 20 seconds under quasi-steady state conditions, ensuring consistent and reliable data collection. This approach allowed for the establishment of a stable aerosol output and provided sufficient time for accurate sampling and analysis. The laser beam crossed the aerosol flow at a right angle and the samplings of the droplet size distribution were performed 10 cm from the atomizer. The scattered signal of the laser beam was recorded with a frequency of 1 Hz, allowing for averaging 20-30 droplet size distribution profiles for each operating condition. More details of the spray measurement setup are given in (*22*).

Although the experiments were conducted using a BUCHI nozzle, its atomization characteristics are consistent with standard pneumatic atomizers such as the Schlick two-fluid nozzle (*11*), which served as the reference geometry for constructing the computational model in this study. Since experimental access to a Schlick nozzle under identical operating conditions was not available, we adopt the BUCHI B-290 droplet size distribution as a representative pneumatic-atomizer distribution; accordingly, our analysis emphasizes the relative influence of gas dynamics on freezing rather than exact nozzle-specific statistics.

## 3. Results and Discussion

### 3.1. *Low temperature in atomizing gas and supercooling of liquid droplets*

We show the trajectories of 50 representative liquid droplets (Figure 5a-c) out of all droplets (total number of 1000) modeled through DPM from the time of being released. The droplets were driven by the swirl jet of the atomizing gas produced by the swirl nozzle, resulting in a swirl pattern in the trajectories of droplets. We color-coded the trajectories with the ambient gas temperature, $T_g$, along the droplets' paths.

We find that most of the droplets shown experienced very low gas temperature, as low as 130 K (Figure 5e), when passing through the low temperature region downstream of the exit of the swirl nozzle (Figure 5d), resulting in significant temperature drop and supercooling in some small droplets of diameter $\sim 1\,\mu$m, within which the droplet temperature, $T_d$, can be as low as 195 K (Figure 5f). The temperature drop in large droplets was smaller because of their larger heat capacity. The cooled droplets warmed up again after leaving the low-temperature region. More details on the flow field are provided in the Supplementary Information.

### 3.2. *Ice nucleation and crystallization in liquid droplets*

To present the ice nucleation and ice crystallization in liquid droplets in detail, we chose five representative droplets, P1–P5 (Figure 6a), from a swirl nozzle as an example, with initial diameters from 1 to 1.6 $\mu$m, which are small enough to freeze. The atomizing gas gauge pressure is set at 5 bar.

The trajectories of the droplets, colored by droplet temperature, show a swirl pattern driven by the swirl flow generated by the swirl channels in the nozzle (Figure 2e). The droplet temperature plummets (Figure 6b) as it passes through the low gas tempera-ture region (Figure 5d), mainly caused by the adiabatic expansion of the pressurized nitrogen. Droplets P1, P3, and P5 reach lower temperatures than P2 and P4. As a result, homogeneous nucleation (dashed lines, Figure 6c), which requires a lower temperature threshold, only occurs in P1, P3, and P5; heterogeneous nucleation (solid lines, Figure 6c), which has a higher temperature threshold, occurs in all droplets. Because P5 reaches the lowest droplet temperature, it has the highest nucleation rates (both homogeneous and heterogeneous nucleation) and has the highest cumulative nucleation of both types (Figure 6d), despite its smallest droplet volume. Results for all 1000 droplets are provided in the Supplementary Information.

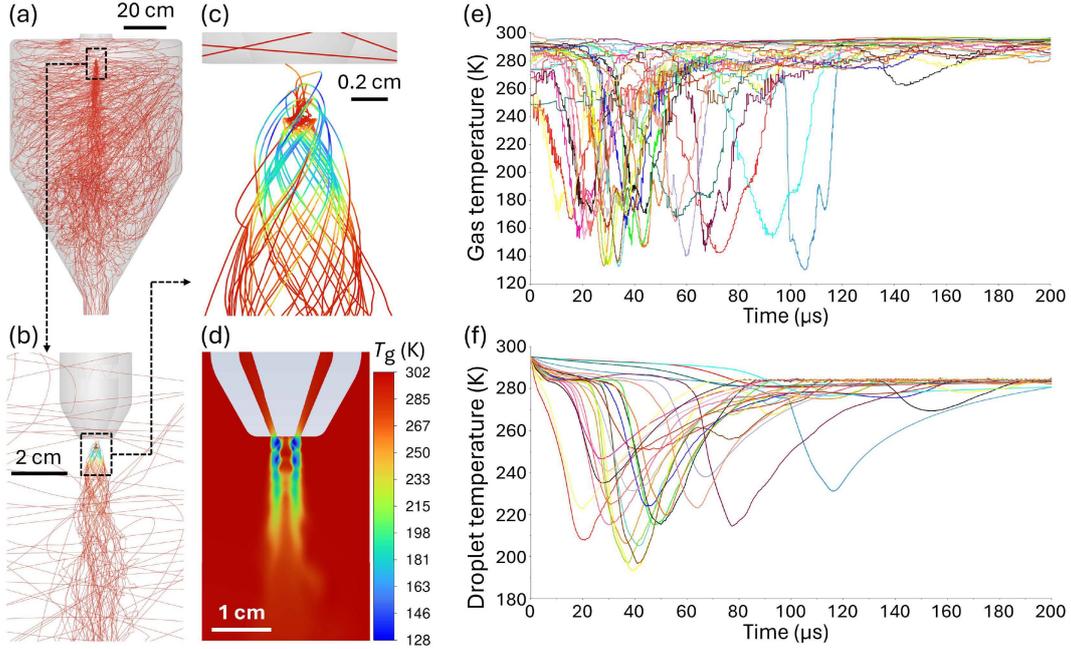

**Figure 5.** Low temperature of the atomizing gas after exiting the nozzle with swirl and supercooling of liquid droplets. The atomizing gas has an initial temperature of 22 °C and an initial pressure of 5 bar (gauge). Overall view (a) and zoomed-in views (b, c) of the trajectories of 50 representative liquid droplets, color-coded by gas temperature, $T_g$. (d) Gas temperature distribution of the atomizing gas exiting the swirl nozzle. (e) Time history of gas temperature, $T_g$, experienced by the droplets during their travel through the chamber. (f) Time history of droplet temperature, $T_d$, during their travel through the chamber.

Ice crystals in supercooled droplets started to grow on ice nuclei as soon as they formed. We model the ice to grow uniformly along the radius, forming a sphere with the ice nucleus at the center. The volume percentage of ice reached 100% for all five droplets (Figure 6e). To compare the speed of ice formation, we compare the ice formation time $\tau$ for the droplets (Figure 6f), where $\tau = t - t_0$, $t_0$ is the starting time of ice formation in each droplet. We find that P5 reached 100% ice within $2\,\mu$s, while it took P2 and P4 about $4\,\mu$s to reach 100% ice. This is a combined effect of a higher number of ice nuclei, lower droplet temperature (affects ice growth speed in Figure 1b), and smaller droplet volume.

### 3.3. *Effect of swirl of atomizing gas on ice formation*

We investigated the effect of swirl of the atomizing gas on ice formation by comparing the case using the nozzle with swirl channels (swirl nozzle, Figure 2e) to the case using a nozzle with straight channels (non-swirl nozzle, Figure 2f). Compared to the swirling droplet trajectories in the swirl nozzle (Figure 5a-c), the droplets of the non-swirl nozzle undergo near rectilinear trajectories after being released (Figure 7a-c). The low gas temperature region of $T_g < 273\,\text{K}$ created by the non-swirl nozzle (Figure 7d) is elongated and extends further than that of the swirl nozzle (Figure 5d). However, the swirl nozzle created larger regions of low gas temperature with $T_g < 160\,\text{K}$, whereas no region of such low temperature is found for the non-swirl nozzle. As a result, both the gas temperature along the droplet trajectory and the droplet temperature of the non-swirl nozzle (Figure 7e,f) had minimum temperatures $T_g$ as low as $190\,\text{K}$ and $T_d$ as low as $220\,\text{K}$, which are higher than those for the swirl nozzle (Figure 5e,f). In addition, because of the elongated low temperature area and its overlap with the near straight droplet trajectory downstream the nozzle, the droplet temperature stayed low for a longer time in the non-swirl case, shown as wide valleys (Figure 7f), than those of the swirl case, shown as sharp valleys (Figure 5f). Swirl creates more extreme minimum temperatures but shorter exposure times in the cold zone, whereas non-swirl yields milder cooling but significantly longer exposure times. More details on the flow field are provided in the Supplementary Information.

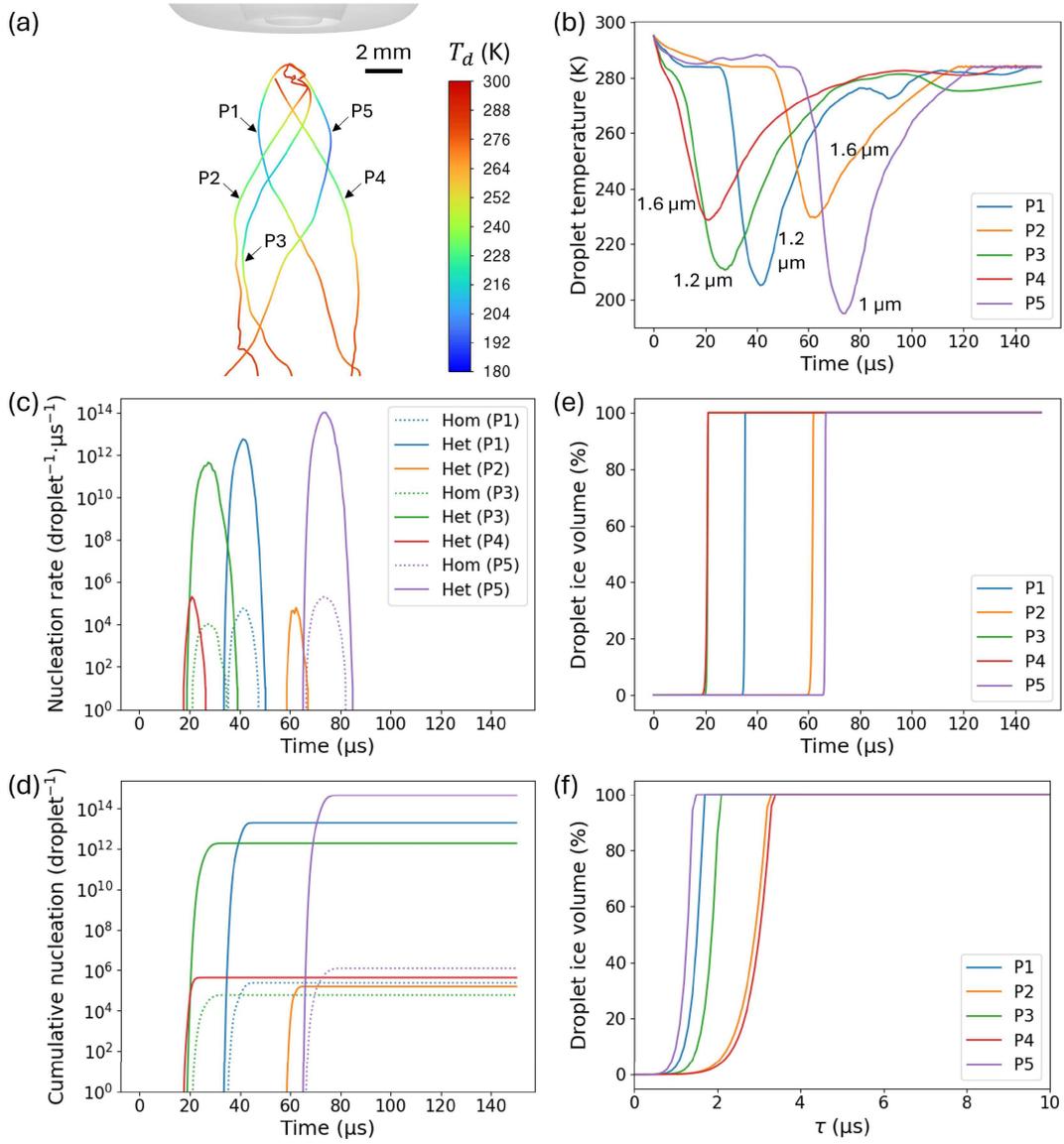

**Figure 6.** Supercooling, ice nucleation, and crystallization of five representative liquid droplets from a swirl nozzle. The atomizing gas has an initial temperature of 22 °C and an initial pressure of 5 bar (gauge). (a) Swirling trajectories of five representative liquid droplets (P1 to P5), color-coded by the droplet temperature. (b) Time history of droplet temperature for the five droplets. (c) Homogeneous and heterogeneous nucleation rates per droplet per microsecond of the five drops. (d) Cumulative number of ice nuclei formed in each droplet by homogeneous and heterogeneous nucleation. (e) Time history of the volume fraction of ice formed in the droplets with respect to the total volume of the droplet. (f) Ice volume fraction in sample droplets since the start of ice formation, where $\tau = t - t_0$ and $t_0$ is the starting time of ice formation in each droplet.

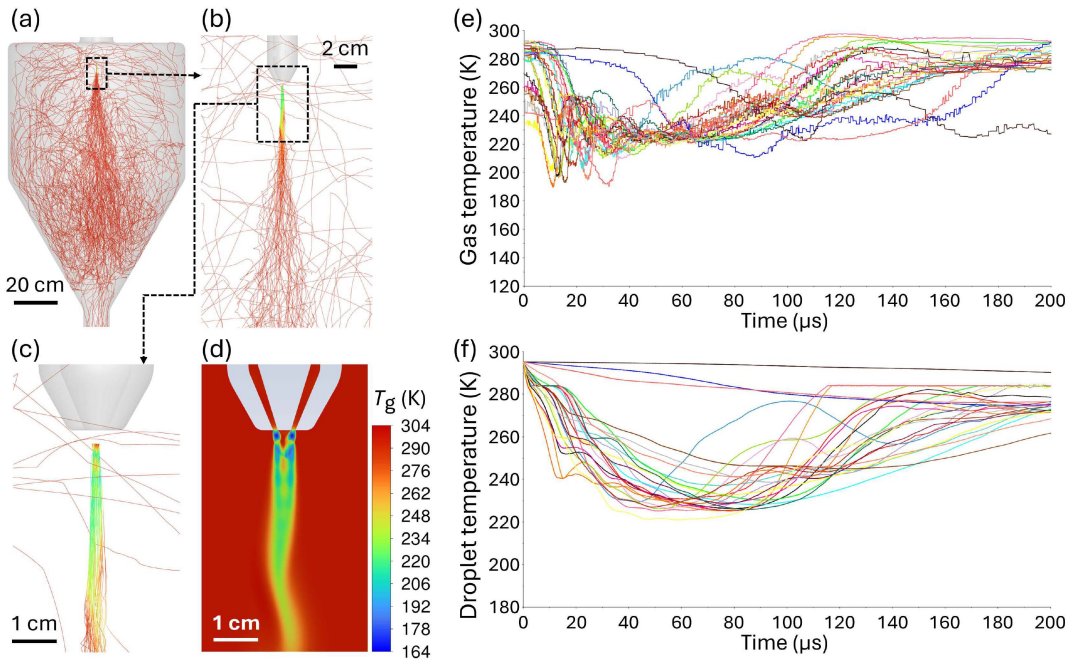

**Figure 7.** Low temperature of the atomizing gas after exiting the nozzle without swirl and supercooling of liquid droplets. The atomizing gas has an initial temperature of 22 °C and an initial pressure of 5 bar (gauge). (a) Overall view and (b, c) zoomed-in views of the trajectories of 50 representative liquid droplets, color-coded by gas temperature. (d) Gas temperature distribution of the atomizing gas exiting the non-swirl nozzle. (e) Time history of gas temperature experienced by the droplets during their travel through the chamber. (f) Time history of droplet temperature during their travel through the chamber.

We chose five representative droplets, P1–P5 (Figure 8a) from the non-swirl nozzle as an example with initial diameters from 1.2 to 1.8 μm. Except for the angle of the 6 channels in the nozzle, all other features are identical between the swirl nozzle and the non-swirl nozzle cases.

The trajectories of the droplets show mostly rectilinear paths driven by the jet of the non-swirl nozzle (Figure 2f). The temperature of the droplets decreased approxi-mately 50 K within 60 µs (Figure 8b). Droplets P3 and P5 reached lower temperatures than the others; therefore, P3 and P5 had a higher heterogeneous ice nucleation rate (Figure 8c). Droplet P4 is 5 to 10 K warmer than P3 and P5 at the low temperature valley. This difference resulted in P4's ice nucleation rate being 3 to 4 orders of magnitude lower than that of P3 and P5. For P1 and P2, which are even warmer than P4, no heterogeneous nucleation occurred (Figure 8c). Therefore, ice growth only happened in P3, P4, and P5, with P3 and P4 reaching 100% ice volume in 6 µs, about 2 µs quicker than P5. For all five droplets chosen, the droplet temperatures were not low enough to trigger homogeneous nucleation. Results for all 1000 droplets are provided in the Supplementary Information.

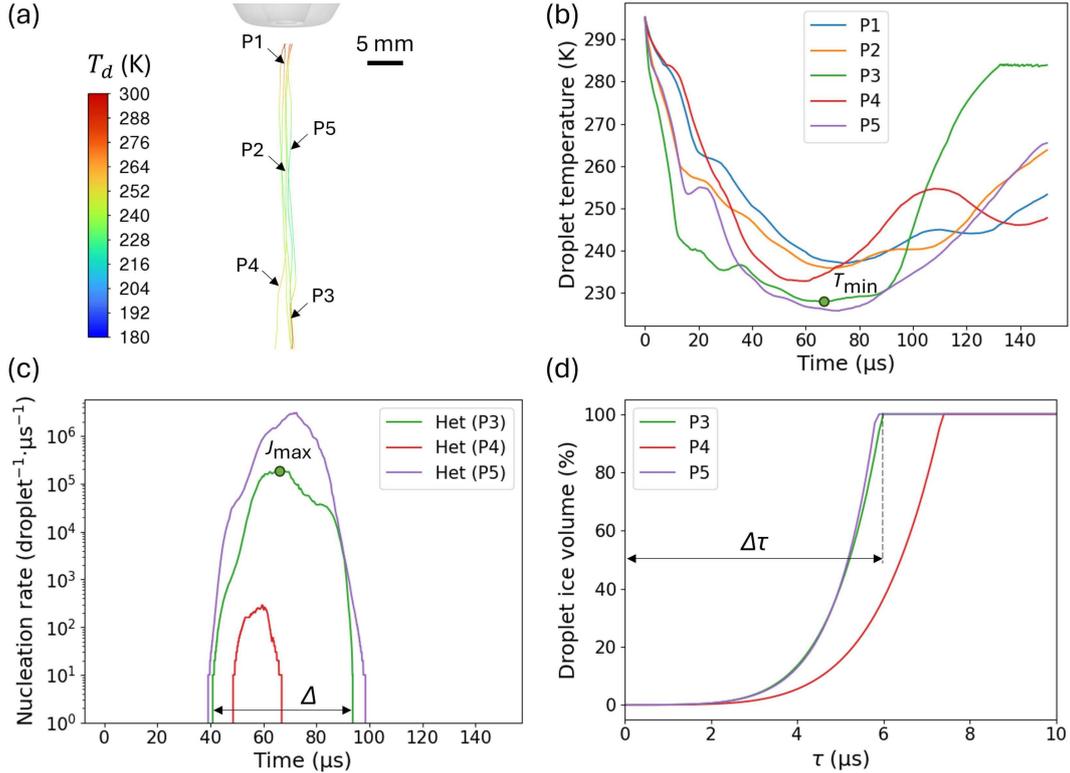

Figure 8. Supercooling, ice nucleation, and crystallization of five representative liquid droplets from the nozzle without swirl. The atomizing gas has an initial temperature of 22 °C and an initial pressure of 5 bar (gauge). (a) Nearly straight trajectories of five representative liquid droplets (P1 to P5), color-coded by the droplet temperature. (b) Time history of droplet temperature for the five droplets. $T_{\min}$ denotes the minimum droplet temperature during this time period. (c) Homogeneous and heterogeneous nucleation rates per droplet per microsecond of the five drops. $\Delta$ denotes the time duration when the nucleation rate is above 1 per droplet per microsecond. (d) Ice volume fraction in droplet since the start point of ice formation, where $\tau = t - t_0$, $t_0$ is the starting time of ice formation in each droplet. $\Delta \tau$ denotes the time needed to reach 100% ice for each droplet.

To quantify the difference in temperature, nucleation rate, and ice growth speed, we define the minimum droplet temperature, $T_{min}$, the maximum nucleation rate, $J_{max}$, and the time to reach 100% ice volume, $\Delta \tau$, and compare them between the swirl nozzle and non-swirl nozzle cases (Figure 8). We first compare the minimum droplet temperature between swirl and non-swirl cases (Figure 9a). For each droplet, we find the $T_{min}$ and $\Delta$, defined as the time duration when the droplet temperature is below 273 K, and we plot the results of all 1000 droplets. We find that most droplets in the swirl case clustered around $\Delta = 50\,\mu$s, while the droplets in the non-swirl case had longer $\Delta$ between 100 and 250 $\mu$s. It is also clear that about half of the droplets in the swirl case reached $T_{min} < 220$ K, while most of the droplets in the non-swirl case had $T_{min} > 220$ K.

We find homogeneous nucleation happened in a distinct number of droplets between the two nozzle configurations: 455 droplets of the swirl case and in only 61 droplets of the non-swirl case (Figure 9b,c). In contrast, heterogeneous nucleation occurred in a comparable number of droplets for both configurations, with 683 droplets in the swirl case and 700 droplets in the non-swirl case. While the maximum heterogeneous nucleation rates of the swirl case are 5 to 6 orders of magnitude higher than the non-swirl case, the time during $\Delta$, defined as the time when the nucleation rate is above 1, of the non-swirl case is around 3 times that of the swirl case, increasing the chance of nucleation and crystallization. We find that droplets in the swirl case reached 100% ice volume in about 2 to 3 $\mu$s, while it took 5 to 7 $\mu$s for those in the non-swirl case to reach 100% ice volume (Figure 9d).

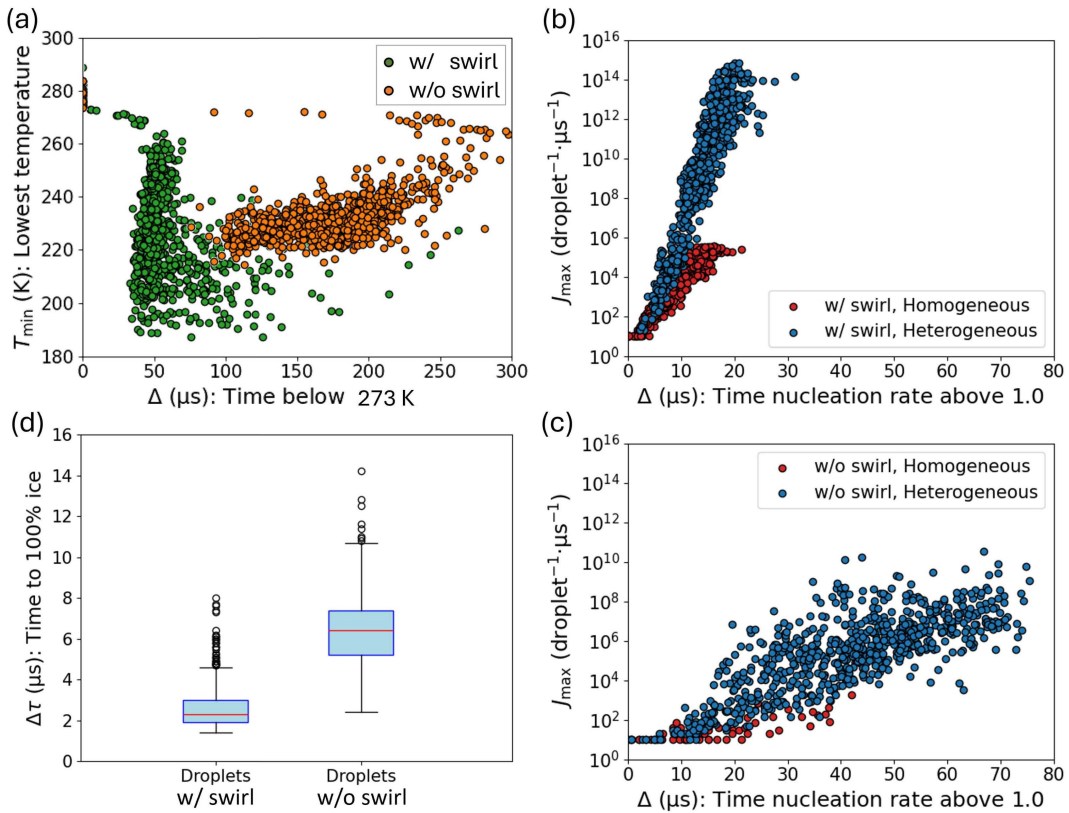

Figure 9. Statistics of droplet temperature and ice nucleation and crystallization. The atomizing gas has an initial temperature of 22 °C and an initial pressure of 5 bar (gauge). (a) Comparison of the lowest droplet temperature between the nozzle with and without swirl. (b) Comparison of the maximum nucleation rate in the swirl nozzle between homogeneous and heterogeneous nucleation. (c) Comparison of the maximum nucleation rate in the non-swirl nozzle between homogeneous and heterogeneous nucleation. (d) Box plot comparison of the time needed to reach 100% ice for each droplet between the nozzle with and without swirl.

To determine the effect of droplet size on ice formation, we classify the 1000 droplets into three categories at each initial droplet diameter: i) iced droplets with both homogeneous and heterogeneous nucleation (100% ice, Hom+Het), ii) iced droplets with only heterogeneous nucleation (100% ice, Het only), and iii) water droplets with no ice formed (zero ice, Figure 10). Note that the droplets have either 100% ice or zero ice based on our calculations. We define $D_{ice}^{100\%}$ as the critical initial droplet diameter below which all droplets fully freeze, whereas $D_{ice}^{0\%}$ denotes the critical diameter above which no droplets freeze. For both the swirl nozzle and non-swirl nozzle using atomizing gas at 5 bar (gauge) and 22 °C, the histograms show that all droplets became ice with a diameter smaller than 1.5 µm ($D_{ice}^{100\%} \approx 1.5\,\mu m$), and no ice was formed in droplets with a diameter larger than 3 µm ($D_{ice}^{0\%} \approx 3\,\mu m$). We also found that the larger the diameter of the droplet, the smaller the proportion of droplets with homogeneous nucleation was. While the number distribution of iced droplets is comparable between the two nozzles, the swirl nozzle (Figure 10a) produced more iced droplets with homogeneous nucleation at small droplet sizes (1.0, 1.2, 1.4, and 1.6 µm) and the non-swirl nozzle (Figure 10b) produced more iced particles at small to median droplet diameter sizes (1.9, 2.2, and 2.5 µm).

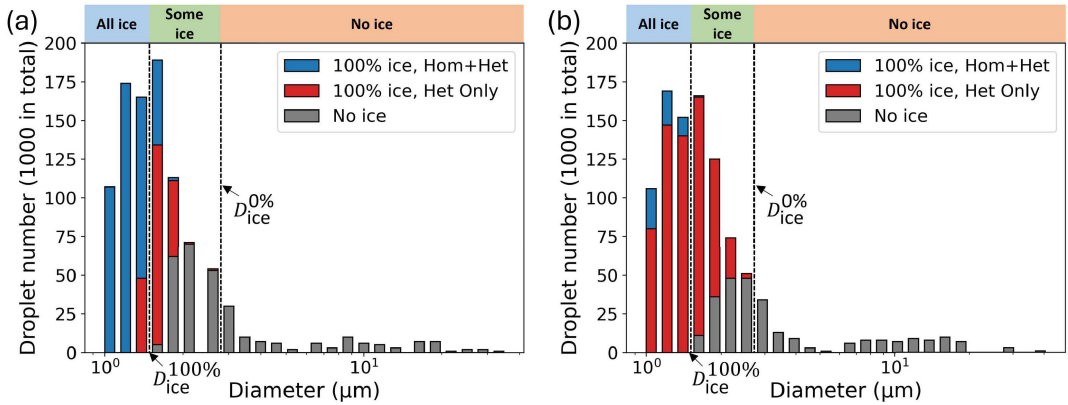

Figure 10. Histogram of the initial diameter of droplets from (a) the nozzle with swirl and (b) the nozzle without swirl. The atomizing gas has an initial temperature of 22 °C and an initial pressure of 5 bar (gauge). Iced droplets that have both homogeneous and heterogeneous nucleation are highlighted in blue. Iced droplets that have only heterogeneous nucleation are shown in red. Water droplets that did not turn into ice are shown in gray. $D_{ice}^{100\%}$ denotes the critical initial droplet diameter below which all droplets fully freeze, whereas $D_{ice}^{0\%}$ denotes the critical diameter above which no droplets freeze.

Despite the difference in diameter distribution of iced droplets between swirl and non-swirl nozzles, both the total number and total volume of iced droplets are close (Table 1). Note that because the iced droplets are small in diameter, although about 70% of the total liquid droplets released (1000 droplets) have turned into iced droplets, they make up only 0.8% of the total volume of the liquid that was released.

Table 1. Comparison of number-based and volume-based iced droplets formation for nozzles with and without swirl channels. The atomizing gas has an initial temperature of 22 °C and an initial pressure of 5 bar (gauge).

|  | Number % of iced droplets | Volume % of iced droplets |
| --- | --- | --- |
| Nozzle with swirl | 68% | 0.8% |
| Nozzle without swirl | 70% | 0.8% |

### 3.4. *Effect of atomizing gas flow rate on crystallization*

To study the effect of atomizing gas flow rate on the icing of droplets, we compare the gas temperature (Figure 11) and the percentage of ice (Figure 12) at atomizing gas gauge pressures of 1.2, 2, 3, 4, and 5 bar, which are within typical operating conditions for two-fluid nozzles. The corresponding atomizing gas flow rates are 0.17, 0.25, 0.34, 0.42, and 0.5 kg/min, and the corresponding gas-to-liquid mass ratio (GLR), here defined as the mass flow rate between the atomizing gas and the liquid feedstock, are 8.4, 12.5, 16.8, 21.0, and 25.1. For all cases in this section, the mass flow rate of liquid is fixed at 0.02 kg/min, and the drying gas flow rate is also fixed, which is the same as all simulation cases in this study. Note that for conciseness, the results of 2 bar are not shown in this section but used in the next section.

We find that the temperature of the drying gas jet decreases and the area of the low temperature region increases as the atomizing gas flow rate increases (Figure 11a-d). The cone angle of the cone-shaped envelope of the droplet trajectories becomes smaller as the flow rate increases and the droplets experience lower temperature (Figure 11e-h).

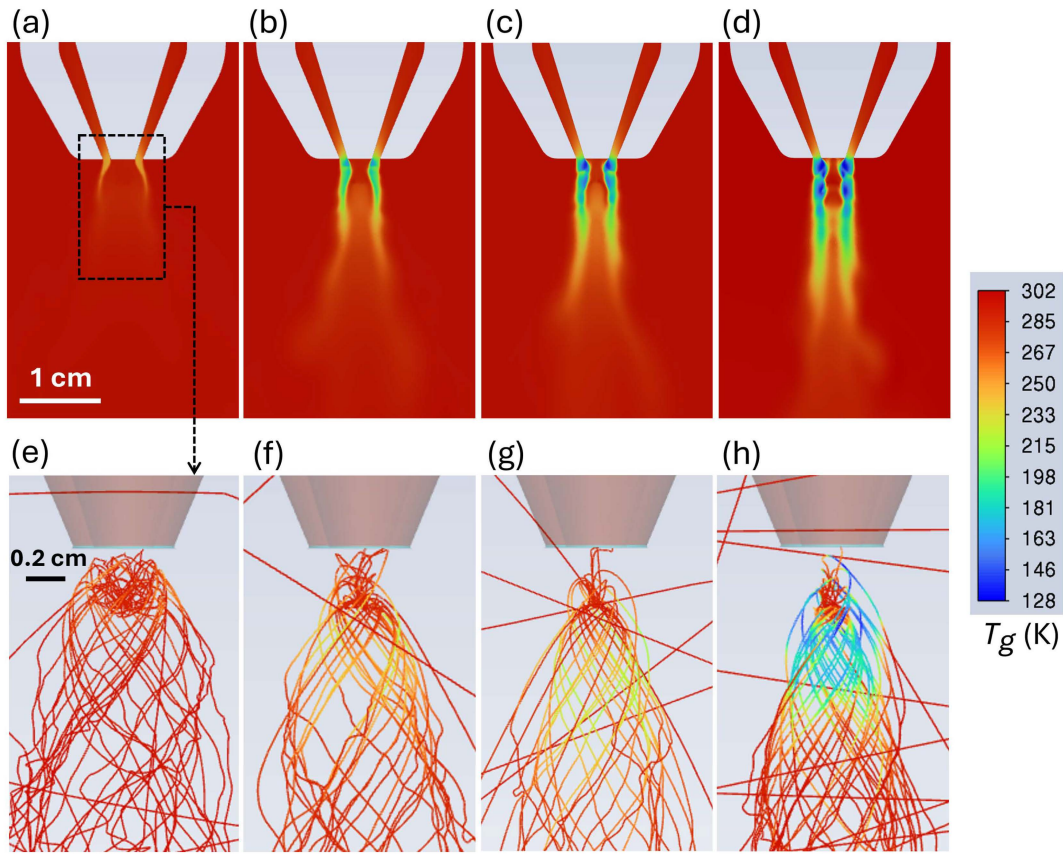

**Figure 11.** Gas temperature for atomizing gas pressure at (a) 1.2 bar, (b) 3 bar, (c) 4 bar, and (d) 5 bar. Liquid droplet trajectories colored by the droplet temperature for atomizing gas gauge pressure at (e) 1.2 bar, (f) 3 bar, (g) 4 bar, and (h) 5 bar. Note that the pressures here are gauge pressures, and the atomizing gas has an initial temperature of 22 °C for all four cases.

We compare the number percentage of iced droplets (Figure 12a) and volume percentage of iced droplets (Figure 12b) between swirl and non-swirl nozzles at gas-to-liquid mass ratios of 8.4, 16.8, 21.0, and 25.1. The number-based percentage and volume-based percentage of iced droplets share a similar trend: They all increase with GLR, and more ice is formed using the swirl nozzle at intermittent GLR. No ice is formed at low GLR for both nozzles. Specifically, ice formed using the swirl nozzle is about four times the ice formed using the non-swirl nozzle at GLR = 21.

Thus increasing GLR, which increases the atomizing gas flow relative to liquid,

intensifies cooling by adiabatic expansion and increases the fraction of droplets that undergo freezing.

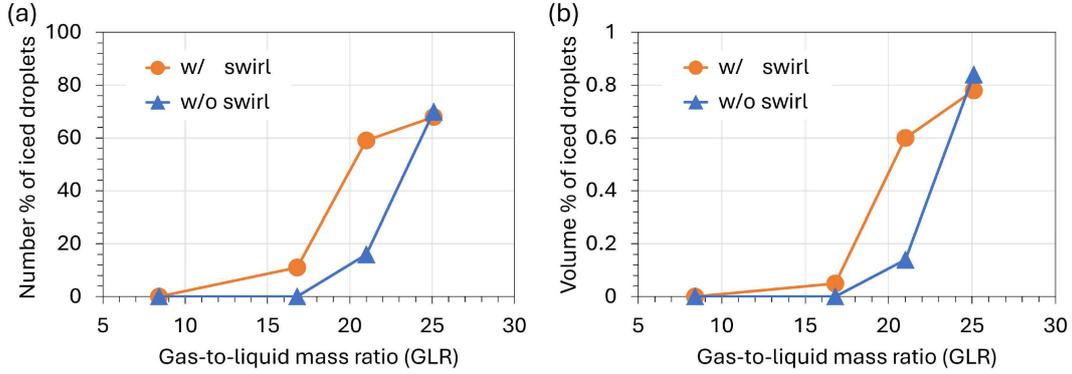

Figure 12. Comparison of (a) the number percentage of iced droplets in all droplets and (b) the volume percentage of ice in all liquid between the nozzle with and without swirl. The atomizing gas has an initial temperature of 22 °C for all cases.

### 3.5. *Effect of atomizing gas temperature on crystallization*

From Table 1, we observe that the ice crystal fraction—expressed in both number and volume percentages—exhibits negligible differences between swirl and non-swirl nozzle configurations. Given that swirl nozzles, particularly Schlick and BUCHI designs, represent the dominant two-fluid atomization technology in industrial applications, we focus our parametric investigation on swirl nozzles. Consequently, the results presented hereafter pertain exclusively to swirl nozzle geometries. This selection is justified by the combination of comparable icing behavior across nozzle types and the practical prevalence of swirl designs in spray systems, enabling us to concentrate on the influence of atomizing gas temperature, pressure and GLR variations in Section 3.5 and 3.6.

To study the effect of atomizing gas temperature on icing, we calculated the ice formation for different gas-to-liquid mass ratios at atomizing gas temperatures of 295 K, 323 K, 353 K, 383 K, and 413 K using the swirl nozzle (Figure 13). For each temperature, we perform simulations at four GLRs of 25.1, 21.0, 16.8, and 12.5, corresponding to atomizing gas gauge pressures of 5 bar, 4 bar, 3 bar, and 2 bar, respectively. The ice percentage decreases as the atomizing gas temperature increases. No ice was found above 383 K (110 °C) for all GLRs, nor for GLR = 12.5 across all atomizing gas temperatures. This suggests that, for the present nozzle and dryer geometry, atomizing gas temperatures ≥110 °C are sufficient to suppress freezing even at high GLR.

We compare the gas temperature downstream of the nozzle (Figure 14a) and the droplet trajectories (Figure 14b) between atomizing gas temperatures of 295 K and 323 K for the swirl nozzle. The patterns of gas jet and droplet trajectories are similar, while the area of the low-temperature region decreases as the atomizing gas temperature increases.

### 3.6. *Operation map to avoid ice formation in droplets*

We plot the heat map of ice percentage with respect to the atomizing gas temperature and GLR for the swirl nozzle (Figure 15). The white dashed line, $P_{\text{no}}(T)$, denotes the boundary of zero ice, whereas ice formation increases towards the top left direction, which is high in GLR and low in atomizing gas temperature. The zero-ice boundary is drawn with piecewise lines connecting the midpoints between the centers of adjacent blocks with and without ice. We obtained the following equation describing the boundary of the ice formation region (white dashed line in Figure 15): $P_{\text{no}}(T) = 3.5$ for $280 \leq T \leq 323$ and $P_{\text{no}}(T) = 3.5 + \frac{1}{30}(T - 323), 323 < T \leq 428$, where $P_{\text{no}}$ is atomizing gas pressure in bar absolute and $T$ is atomizing gas temperature in K.

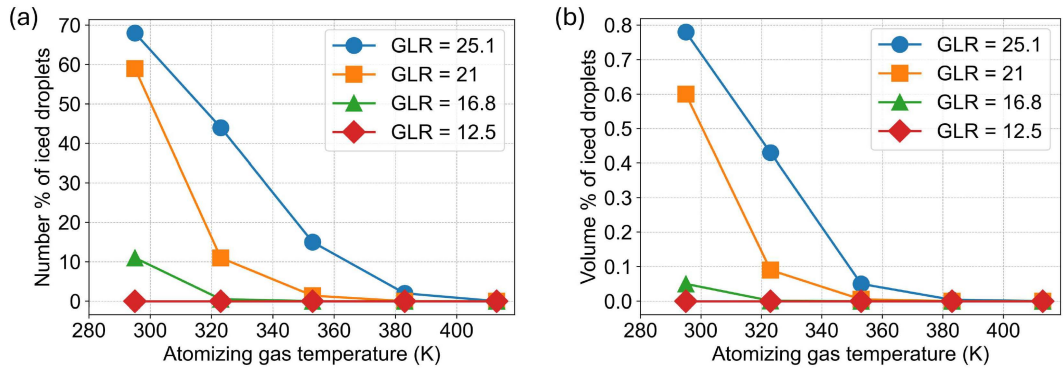

**Figure 13.** (a) Number percentage of iced droplets in all droplets and (b) volume percentage of ice in sprayed liquid for swirl nozzle with respect to the atomizing gas temperature for gas-to-liquid mass ratios (GLR) of 25.1, 21.0, 16.8, and 12.5, corresponding to atomizing gas gauge pressures of 5 bar, 4 bar, 3 bar, and 2 bar, respectively. Note that all pressures here are gauge pressures.

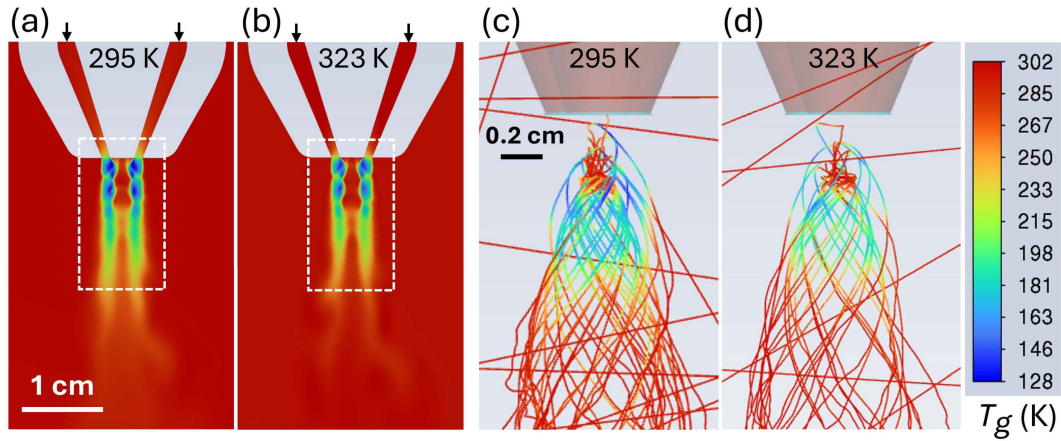

**Figure 14.** Gas temperature (a and b) and trajectory of droplets colored by gas temperature (c and d) at 295 K and 323 K. The atomizing gas has an initial pressure of 5 bar (gauge).

The red solid line, $P_{\text{adi}}(T)$, in Figure 15 is based on the assumption of the ideal adiabatic expansion process for the atomizing gas expanding from the absolute stagnation pressure of 6 bar (5 bar gauge) and stagnation temperature of 383 K, which is the first no ice condition we found for the GLR = 25.1 case (see heat maps in Figure 15). The adiabatic expansion relation is

$$P_{\text{adi}}(T) = P_0 \left(\frac{T}{T_0}\right)^{\gamma/(\gamma-1)}, \qquad (280 < T < 428), \tag{11}$$

where $P_{\text{adi}}(T)$ is in bar absolute and $T$ is in K, and $\gamma$ is the specific heat ratio of nitrogen (taken as $\gamma = 1.4$). The pair $(P_0, T_0)$ denotes the absolute stagnation pressure and stagnation temperature of the atomizing gas at a no ice condition adjacent to an ice condition. Thus, $P_{\text{adi}}(T) = 6 \left(\frac{T}{383}\right)^{3.5}$, where $P_{\text{adi}}$ is in bar absolute and $T$ is in K.

From Figure 15, it can be seen that the zero-ice boundary (white dashed line, $P_{\text{no}}(T)$) and the ideal adiabatic expansion curve (red line, $P_{\text{adi}}(T)$) exhibit similar trends and closely matching values in the atomizing gas temperature range $323 < T \leq 428$ K. This indicates that the ideal adiabatic expansion process provides a reasonable approximation for estimating ice formation in the swirl nozzle setup.

Accordingly, we propose that the ideal adiabatic expansion curve, $P_{\text{adi}}(T)$, serves as a physically grounded predictor for the ice-formation boundary in the operational map. The stagnation conditions $(P_0, T_0)$ vary with nozzle type (swirl versus non-swirl) and design parameters including nozzle internal geometry (channel length, twist angle, hydraulic diameter, discharge coefficient), internal pressure losses and viscous dissipation (typically more pronounced in swirl channels), upstream line and regulator pressure drops, gas preheating and thermal losses to the nozzle body, and gas composition (through $\gamma$).

To ensure ice-free operation, we propose a conservative guideline based on shifting the ideal adiabatic expansion curve toward higher temperatures by $\Delta T = 20$ K. This rightward shift accounts for non-ideal effects (heat losses, local over-expansion, and spatial temperature non-uniformity) and ensures the safe curve envelopes the fitted no-ice boundary $P_{\text{no}}(T)$. The safe operating curve is defined as

$$P_{\text{safe}}(T) = P_{\text{adi}}(T - \Delta T) = P_0 \left(\frac{T - \Delta T}{T_0}\right)^{\gamma/(\gamma-1)}, \qquad (280 < T < 428), \tag{12}$$

where $P_{\text{safe}}$ is in bar absolute and $T$ is in K. In this study, we established $P_0 = 6$ bar, $T_0 = 383$ K and $\Delta T = 20$ K for the investigated two-fluid swirl nozzle design and nitrogen–water atomization system.

Operating on the no-ice side of this curve—at sufficiently high $T$ for a given $P$ (or equivalently, sufficiently low $P$ for a given $T$)—should remain ice-free, with $P_{\text{safe}}(T)$ serving as a conservative upper bound. Operators can thus use $P_{\text{safe}}$ as a practical limit: keeping the atomizing gas pressure below $P_{\text{safe}}$ at the chosen temperature predicts ice-free spray.

This conservative guideline applies specifically to the two-fluid nitrogen–water system and swirl nozzle design investigated. Significant changes to nozzle design, gas composition, or dryer configuration require re-evaluation of $P_{\text{safe}}(T)$.

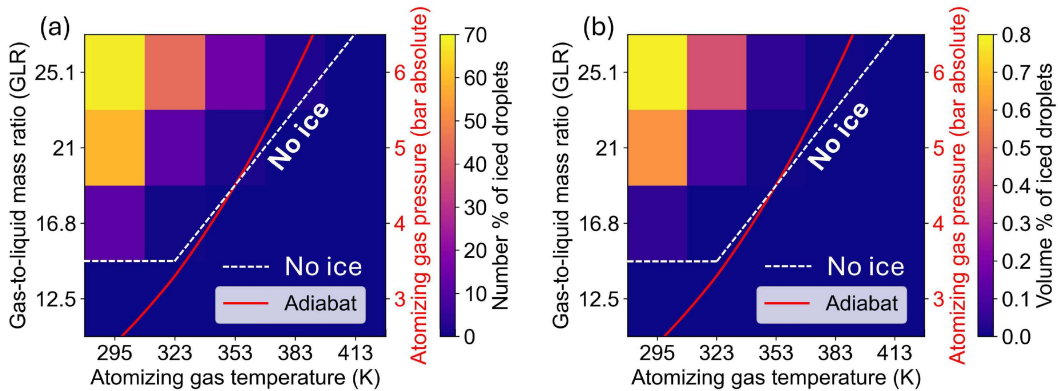

Figure 15. Heat maps for (a) the number fraction of iced droplets in the spray and (b) the volume fraction of ice in the sprayed liquid for the swirl nozzle, as functions of atomizing gas temperature and gas-to-liquid mass ratio (GLR). Results are shown for GLR values of 25.1, 21.0, 16.8, and 12.5, corresponding to absolute atomizing gas pressures of 6, 5, 4, and 3 bar, respectively. The red curve, $P_{\text{adi}}(T)$, denotes the ideal adiabatic expansion trajectory of the atomizing gas for a stagnation condition of 6 bar absolute and 383 K. The white dashed curve, $P_{\text{no}}(T)$, marks the zero-ice boundary, constructed by connecting the midpoints between the centers of adjacent bins with and without ice.

Although the detailed GLR–pressure–temperature phase map and the corresponding $P_{\text{safe}}(T)$ relation in Eq. (12) were constructed for the swirl nozzle configuration, the non-swirl nozzle systematically exhibits higher minimum gas and droplet temperatures at the same operating conditions, while producing comparable overall number and volume fractions of iced droplets. Taken together, these observations indicate that the no-ice boundary identified for the swirl nozzle provides a conservative criterion for non-swirl two-fluid nozzles using nitrogen: for a given atomizing gas temperature, non-swirl configurations operated at or below the pressure predicted by $P_{\text{safe}}(T)$ in Figure 15 are also expected to remain free of flash freeze–thaw of droplets, and may in practice tolerate slightly higher pressures before droplet freezing occurs.

The key implication from a pharmaceutical/biopharmaceutical perspective is that a sizable fraction of the smallest droplets, which represent only ∼1% of the spray volume but a much larger fraction of the interfacial area due to their large number fraction (see Figures 12, 13, and 15), may undergo rapid freeze–thaw cycles on microsecond timescales. Pharmaceutical formulations that concentrate freeze-thaw-sensitive actives into such collection of smallest droplets during spray drying are at high risk of freeze-thaw-induced degradation, even under standard spray drying conditions that are generally considered "warm".

## 4. Conclusions

We combined computational fluid dynamics simulations with analytical modeling to investigate the flow generated by a two-fluid nozzle—an atomizing gas stream transporting liquid droplets, as commonly employed in spray drying operations. The gas expansion at the nozzle exit induces adiabatic cooling of the atomizing gas stream, which in turn cools, and under suitable conditions freezes, the dispersed spray droplets, thereby creating a previously underappreciated *flash freeze–thaw* exposure very close to the nozzle.

At a representative operating condition with atomizing gas at 5 bar (gauge) and spray flow rate of 0.02 kg/min (20 mL/min), the gas temperature at the nozzle outlet can drop to $-130$ °C from 22 °C due to adiabatic expansion. Under these conditions, both homogeneous and heterogeneous ice nucleation can occur in micron-sized supercooled water droplets (diameter < 3 $\mu$m) that traverse the low-temperature regions of the atomizing gas. For an atomizing gas at 5 bar (gauge) and 22 °C, all droplets smaller than 1.5 $\mu$m freeze, whereas droplets larger than 3 $\mu$m remain liquid for both swirl and non-swirl nozzle designs. As many as 70% of the droplets completely freeze within several microseconds before exiting the cold region, after which they rapidly thaw, so that the frozen droplets account for only about 0.8% of the total sprayed-liquid volume. In other words, a large population of the smallest droplets undergoes a complete flash freeze–thaw cycle on a microsecond time scale, even though the overall spray-drying process is operated under nominally warm conditions.

From a pharmaceutical and biopharmaceutical standpoint, this transient flash freezing and thawing of the smallest droplets—despite their modest contribution to volume—can disproportionately affect product quality, because these droplets contribute a large share of the total interfacial area and may be enriched in freeze–thaw-sensitive actives.

Comparison of swirl and non-swirl nozzle designs shows that, in the swirl configuration, droplets experience lower minimum temperatures but for shorter residence times in the cold region, whereas in the non-swirl nozzle they are exposed to somewhat higher minimum temperatures over longer timescales. Despite these differences in thermal history, both nozzle types generate comparable number fractions and volume fractions of frozen droplets.

The parametric analysis, in which atomizing gas temperature and gas-to-liquid mass ratio (GLR) were systematically varied, demonstrates that ice formation in sprayed droplets increases with GLR and decreases with increasing atomizing gas temperature. In particular, for the present nozzle and dryer geometry, raising the atomizing gas inlet temperature above $\sim 110$ °C ($\sim 380$ K) suppresses freezing for all GLRs in the range 8–25 studied here. Alternatively, decreasing the GLR below $\sim 12$ also prevents ice formation in sprayed droplets for the entire range of investigated atomizing gas temperatures, 22–140 °C (295–413 K). Remarkably, variations in the temperature of the bulk drying gas have negligible influence on near-nozzle droplet freezing.

Phase maps constructed from these numerical simulations for the swirl nozzle delineate the operating window in which ice formation—and thus flash freeze–thaw of droplets—is avoided, as given by Figure 15. Moreover, the adiabatic law for an ideal gas provides a physically grounded and convenient indicator for identifying operating conditions that suppress ice formation in the spray. Comparison of the numerically determined no-ice boundary with the ideal-gas adiabatic expansion curve shows that a modest temperature-shifted adiabatic relation, $P_{\text{safe}}(T)$, given by Eq. (12), furnishes a conservative upper bound on atomizing gas pressure for ice-free operation of the nitrogen–water, two-fluid spray swirl nozzle configuration studied here. Finally, because the swirl nozzle produces lower minimum temperatures than the non-swirl case while yielding similar iced-droplet fractions, the no-ice boundary in Figure 15 and Eq. (12) provide a conservative operating criterion for both swirl and non-swirl two-fluid nozzles using nitrogen.

## 5. Acknowledgment

This work was supported by the National Science Foundation (NSF) STTR Phase I award No. 2304461 and NSF SBIR Phase II award No. 2451720, and also supported by the National Institutes of Health STTR Phase I award No. 1R41TR004571-01A1. The numerical simulations were performed on computational resources managed and supported by Princeton Research Computing of the Office of the Dean for Research of Princeton University. The authors gratefully acknowledge Prof. Arun S. Mujumdar (McGill University and Western University, Canada) for his reading of an earlier version of this manuscript and for insightful comments that improved its quality.

## 6. Data Availability Statement

The data that support the findings of this study are available within the article.

## 7. Declaration of Competing Interest

M.M., Z.P., and H.A.S. hold equity in Inaedis, Inc.

SUPPLEMENTARY INFORMATION

# Flash Freeze–Thaw Phenomenon in Sprayed Evaporating Micrometer Droplets

Junshi Wang[a], Zehao Pan[a,b], Howard A. Stone[a,b], and Maksim Mezhericher[a,b]

[a]Department of Mechanical and Aerospace Engineering, Princeton University, Princeton, New Jersey 08544, USA
[b]Inaedis, Inc., Princeton, New Jersey 08540, USA

## 1. Droplet size distribution modeling

To model the initial droplet size distribution in our numerical simulations, we created a user-defined function (UDF, see Table S1) of the tabulated droplet size distribution as input for the droplet injections in the DPM model of ANSYS Fluent. We chose mass fraction as the input for the injections and filled other columns with zeros. The table has a fixed format so that the ANSYS Fluent software can read it correctly. In this table, "drop-diam" is the droplet diameter with the unit in meter (m), "num-frac" is the number-based fraction, "mass-frac" is the volume-based fraction, "cum-num-frac" is the cumulative number-based fraction, and "cum-vol-frac" is the cumulative volume-based fraction. The UDF-modeled droplet size distribution corresponds to our experimental measurement of the actual droplet size distribution produced by the BUCHI two-fluid spray nozzle, as described in the main paper text (see Section 2.5 and Figure 4 of the article).

## 2. Homogeneous and heterogeneous nucleation rates and ice growth

Figure S1 demonstrates the results of numerical modeling of the homogeneous and heterogeneous nucleation rates and ice growth for all the frozen droplets sprayed by nozzles with swirl and without swirl. The atomizing gas has an initial temperature of 22°C and an initial pressure of 5 bar (gauge). Analysis of these plots reveals that droplets of the non-swirl nozzle yielded less homogeneous nucleation and stayed longer in the cold region than those of the swirl nozzle. The droplets took longer to freeze when the non-swirl nozzle was used compared to the case with the swirl nozzle. These findings are consistent with the statistical data presented in the main article text.

---

Corresponding author: Maksim Mezhericher. Email: maksymm@princeton.edu

Table S1. User-defined function (UDF) of droplet size distribution written in ANSYS Fluent tabulated form for the nozzle injections. Here "drop-diam" is the droplet diameter with the unit in meter (m), "num-frac" is the number-based fraction, "mass-frac" is the volume-based fraction, "cum-num-frac" is the cumulative number-based fraction, and "cum-vol-frac" is the cumulative volume-based fraction. We used mass fraction as the input for the injections and filled other columns with zeros, as required by the UDF convention of ANSYS Fluent software.

| drop-diam | num-frac | mass-frac | cum-num-frac | cum-vol-frac |
|---|---|---|---|---|
| $8.6 \times 10^{-7}$ | 0 | 0 | 0 | 0 |
| $1.00 \times 10^{-6}$ | 0 | 0.04 | 0 | 0 |
| $1.17 \times 10^{-6}$ | 0 | 0.10 | 0 | 0 |
| $1.36 \times 10^{-6}$ | 0 | 0.17 | 0 | 0 |
| $1.58 \times 10^{-6}$ | 0 | 0.24 | 0 | 0 |
| $1.85 \times 10^{-6}$ | 0 | 0.28 | 0 | 0 |
| $2.15 \times 10^{-6}$ | 0 | 0.31 | 0 | 0 |
| $2.51 \times 10^{-6}$ | 0 | 0.30 | 0 | 0 |
| $2.93 \times 10^{-6}$ | 0 | 0.27 | 0 | 0 |
| $3.41 \times 10^{-6}$ | 0 | 0.21 | 0 | 0 |
| $3.98 \times 10^{-6}$ | 0 | 0.16 | 0 | 0 |
| $4.64 \times 10^{-6}$ | 0 | 0.14 | 0 | 0 |
| $5.41 \times 10^{-6}$ | 0 | 0.22 | 0 | 0 |
| $6.31 \times 10^{-6}$ | 0 | 0.46 | 0 | 0 |
| $7.36 \times 10^{-6}$ | 0 | 0.95 | 0 | 0 |
| $8.58 \times 10^{-6}$ | 0 | 1.78 | 0 | 0 |
| $1.000 \times 10^{-5}$ | 0 | 2.98 | 0 | 0 |
| $1.166 \times 10^{-5}$ | 0 | 4.54 | 0 | 0 |
| $1.359 \times 10^{-5}$ | 0 | 6.35 | 0 | 0 |
| $1.585 \times 10^{-5}$ | 0 | 8.21 | 0 | 0 |
| $1.848 \times 10^{-5}$ | 0 | 9.85 | 0 | 0 |
| $2.154 \times 10^{-5}$ | 0 | 10.97 | 0 | 0 |
| $2.512 \times 10^{-5}$ | 0 | 11.35 | 0 | 0 |
| $2.929 \times 10^{-5}$ | 0 | 10.86 | 0 | 0 |
| $3.415 \times 10^{-5}$ | 0 | 9.59 | 0 | 0 |
| $3.981 \times 10^{-5}$ | 0 | 7.74 | 0 | 0 |
| $4.642 \times 10^{-5}$ | 0 | 5.64 | 0 | 0 |
| $5.412 \times 10^{-5}$ | 0 | 3.63 | 0 | 0 |
| $6.310 \times 10^{-5}$ | 0 | 1.94 | 0 | 0 |
| $7.356 \times 10^{-5}$ | 0 | 0.73 | 0 | 0 |
| $8.577 \times 10^{-5}$ | 0 | 0.02 | 0 | 0 |

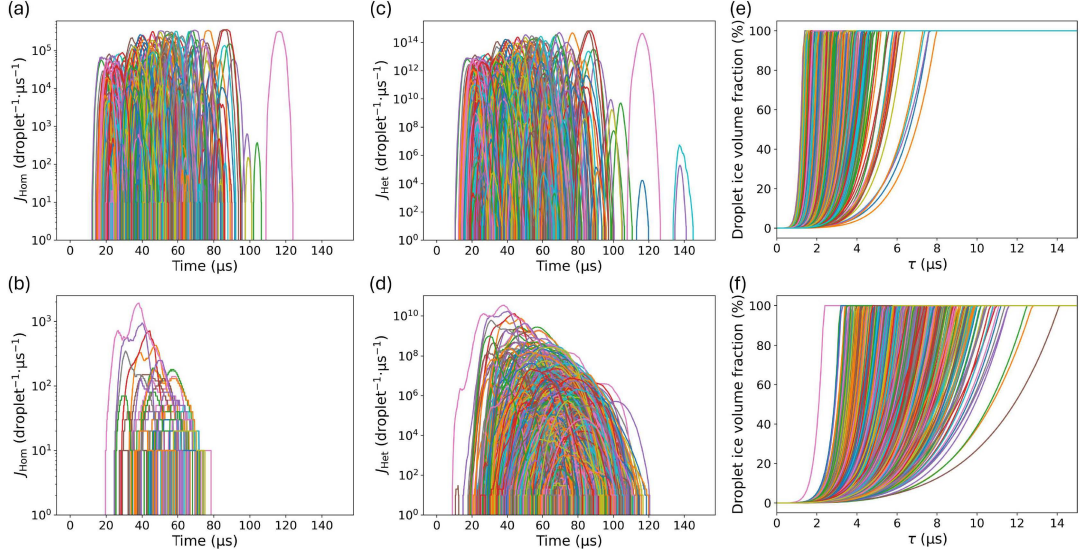

**Figure S1.** Nucleation and ice growth of all the frozen droplets. Homogeneous nucleation rate per droplet per microsecond, $J_{\text{Hom}}$, for all droplets of (a) the nozzle with swirl and (b) the nozzle without swirl. The atomizing gas has an initial temperature of 22°C and an initial pressure of 5 bar (gauge). Each line corresponds to a single droplet, and the different colors are used to distinguish one droplet from another. Heterogeneous nucleation rate per droplet per microsecond, $J_{\text{Het}}$, for all droplets of (c) the nozzle with swirl and (d) the nozzle without swirl. Ice volume fraction in droplet since the start point of ice formation for all droplets of (e) the nozzle with swirl and (f) the nozzle without swirl, where $\tau = t - t_0$, $t_0$ is the starting time of ice formation in each droplet.

## 3. Gas flow fields downstream of a nozzle

Here, we provide more details on the gas flow field information downstream of a nozzle for both the swirl nozzle and the non-swirl nozzle (Figure S2). The atomizing gas has an initial temperature of 22°C and an initial pressure of 5 bar (gauge). We find that the low-temperature pockets downstream of the nozzle (Figure S2e,f) correspond to the large negative values in z-velocity (Figure S2c,d). Correspondingly, these flow fields indicate that both the high-magnitude negative axial velocity and the cooling of the gas in the chamber are driven by the expansion of the atomizing gas upon release from the nozzle exit.

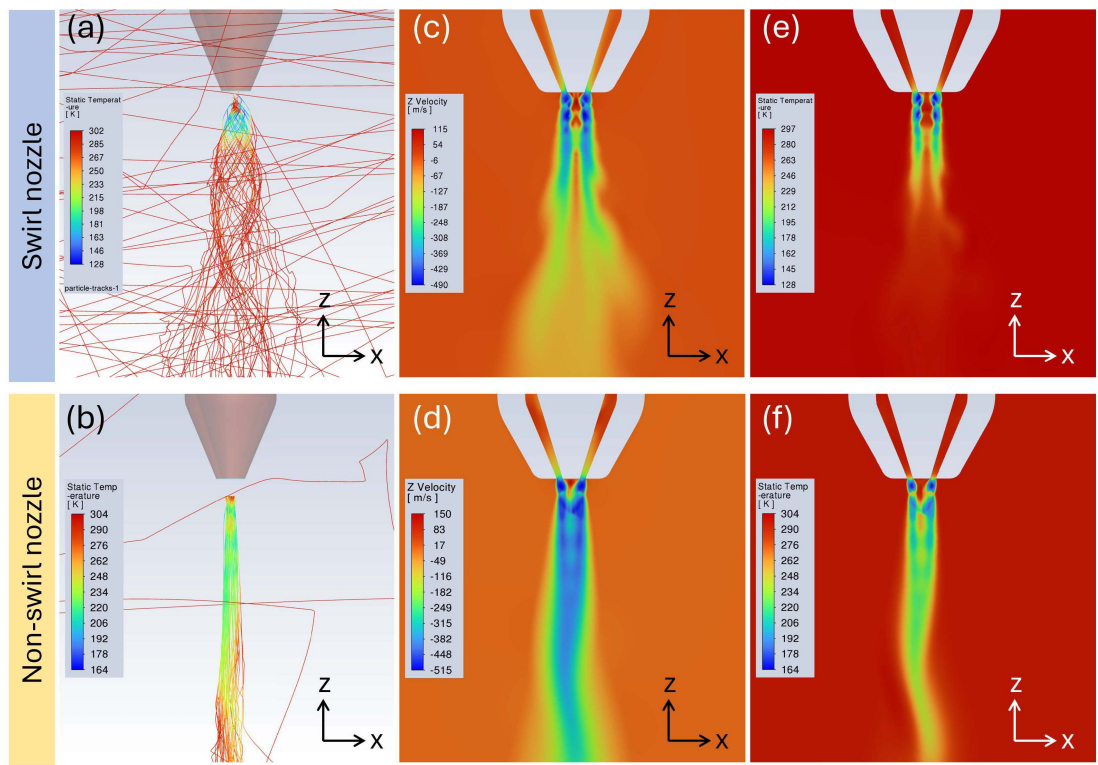

**Figure S2.** Magnified view of the trajectories of 50 representative liquid droplets, color-coded by gas temperature, for (a) nozzle with swirl and (b) nozzle without swirl. The atomizing gas has an initial temperature of 22°C and an initial pressure of 5 bar (gauge). Signed value of the z-component velocity of the atomizing gas along the path of the droplets for (c) nozzle with swirl and (d) nozzle without swirl. Temperature of the atomizing gas along the path of the droplets for (e) nozzle with swirl and (f) nozzle without swirl.

4